\documentstyle[twocolumn,prl,aps]{revtex}
\begin{document}
\draft
\title{Tricritical behavior of the frustrated XY antiferromagnet}
\author{M.L. Plumer, A. Mailhot,\cite{note} and A. Caill\'e}
\address{ Centre de Recherche en Physique du Solide et D\'epartement de
Physique}
\address{Universit\'e de Sherbrooke, Sherbrooke, Qu\'ebec, Canada J1K 2R1}
\date{May 1994}
\maketitle
\begin{abstract}
Extensive histogram Monte-Carlo simulations
of the XY antiferromagnet on a stacked triangular lattice reveal
exponent estimates which strongly favor a scenario of mean-field tricritical
behavior for the spin-order transition.  The corresponding chiral-order
transition occurs at the same temperature but appears to be decoupled
from the spin-order.  These results are relevant to a wide
class of frustrated systems with planar-type order and serve to resolve a
long-standing controversy regarding their criticality.
\end{abstract}

The nature of phase transitions in frustrated systems which can
be mapped onto spin models has been studied extensively over the
past twenty years by means of renormalization-group (RG) methods and
Monte-Carlo (MC) simulations \cite{book}.  The triangular antiferromagnet
serves
as the simplest example of geometry-induced frustration, which in this
case gives rise to the non-colinear $120^\circ$ spin sturcture.  This type of
magnetic order can also be described as a helically polarized spin density.
The appropriate Landau-Ginzburg-Wilson Hamiltonian is usually taken as
\begin{eqnarray}
{\cal H}=r{\bf S} \cdot {\bf S^\ast}
+ {\bf \nabla S} \cdot {\bf \nabla S^\ast}
+ U_1({\bf S} \cdot {\bf S^\ast})^2
+ U_2\mid {\bf S} \cdot {\bf S} \mid ^2,
\nonumber
\end{eqnarray}
where ${\bf S}$ is a complex vector and there are {\it two} fourth-order terms
as a consequence of frustration.
One reason for the plethora of studies of this Hamiltonian
is that in addition to describing
helical spin systems,
it is also relevant to the dipole phase of superfluid
$^3He$, Josephson-junction arrays in a transverse magnetic
field, as well as the fully frustrated bipartite
lattice (see \cite{book} and \cite{kawa1} for references).
Although some of the earlier RG studies suggested a continuous
transition within standard universality classes, others found
evidence for a first-order transition.
In view of these results, the tricritical behavior suggested
by the histogram MC
simulations of the XY stacked triangular antiferromagnet (STAF)
reported here may not be too surprising.

Interest in the criticality of frustrated spin systems has recently
been enhanced due to the suggestion by Kawamura of new ``chiral" universality
classes associated with the XY and Heisenberg STAF's \cite{kawa1,kawa2}.
This claim is partially supported by arguments which demonstrate that the
symmetry of the order parameter $V$ in the XY case
involves a discrete two-fold chiral
degeneracy as well as that of the two-dimensional rotation group, so that
$V=Z_2 \times S_1$ \cite{lee}.  If this is
indeed the relevant symmetry, and if the transition is neither first-order
nor tricritical, then the universality class should be different from the
standard ones.  The strongest support for the existence of new universality
classes comes from Kawamura's MC simulations.  These were of the
conventional type using rather large lattices
$L \times L \times L$ with L=18-60 but with a possibly modest number of Monte
Carlo steps per site (MCS), 6-20 runs with $2 \times 10^4$ MCS each.
Critical exponents were estimated by the conventional
``data collapsing" method, which necessitates a simultaneous estimation of the
critical temperature.  In the Heisenberg case, the reported values are
$\alpha=0.24(8), \beta=0.30(2), \gamma=1.17(7)$ and $\nu=0.59(2)$.  These
results have recently been corroborated by three different groups using
the more accurate finite-size scaling based on histogram MC
data \cite{bhat,loi,mail1}.
In the XY case, the exponent values reported by Kawamura
are $\alpha=0.34(6), \beta=0.253(10), \gamma=1.13(5)$ and $\nu=0.54(2)$.
Both of these sets of exponents are
quite different from those of any standard class in 3D.
However, they are somewhat suggestive of
mean-field tricriticality (apart from the values for $\alpha$), where
$\alpha=\frac12, \beta=\frac14, \gamma=1$ and $\nu=\frac12$.
Note that it is only
Kawamura's estimate for $\beta$ in the XY case which coincides with these
values within error.

Unique sets of exponents were also found for the chirality order, which
Kawamura suggests occurs at the same critical temperature.  The coincidence,
or not, of the two critical temperatures has received much attention in
corresponding 2D frustrated systems \cite{lee,chiral}.
Most authors appear to support the notion that they are the same.

In contrast with the suggestion of new chiral universality,
the conclusion of a study of the non-linear sigma model
in $2+\epsilon$ dimensions by Azaria {\it et al.} \cite{aza} is that the
criticality of such frustrated systems, at least in the Heisenberg case,
is nonuniversal.
Depending on unspecified system parameters, one can have
either standard $O(4)$ criticality, a first-order transition, or mean-field
tricritical behavior.  It is natural to extend these arguments to the XY
case and speculate either a first-order or tricritical
transition \cite{azapri}.  We note that the nature of the chiral transition
was not addressed by these authors.

This already confusing situation was recently further exacerbated by the
results of Zumbach's local-potential-approximation treatment of the
RG\cite{zum}.  This work emphasizes the possibility of
``almost second-order phase transitions" for frustrated systems,
where there can be a set of effective critical exponents
(also see Refs. \onlinecite{kawa1} and \onlinecite{peczak}).
At least in the Heisenberg
case, the distinction between this possibility and that of a new
chiral universality class may never be satisfactorily resolved by MC
simulations.

The finite-size scaling of thermodynamic functions evaluated from
histogram MC data has demonstrated the ability to yield highly accurate
critical-exponent estimates for unfrustrated systems \cite{pec2,ferr}.
In addition, this procedure when combined with the cumulant crossing
method gives an independent and accurate estimate of critical
temperatures \cite{pec2}.
This latter feature is particularly useful for frustrated
systems in view of the possibility that spin and chiral degrees of freedom
order at different, but nearby, temperatures.

It is hoped that the results of MC simulations presented here
will provide convincing evidence that the XY STAF exhibits mean-field
tricritical behavior.

Near-neighbor antiferromagnetic exchange coupling
in the basal plane, $J_\bot = 1$, and ferromagnetic coupling along
the $c$-axis, $J_\| = -1$, were used \cite{plum1}.
The Metropolis MC algorithm was employed in combination with the histogram
technique on lattices with L=12-33 and runs using
$1 \times 10^6$ MCS for the smaller lattices and $1.2 \times 10^6$ MCS for the
larger lattices, after discarding the initial $2 \times 10^5$ - $5 \times 10^5$
MCS for thermalization.  Averaging was then made over 6 (smaller L) to
17 (larger L) runs.  For the largest lattices, this gives a respectable
$20.4 \times 10^6$ MCS for averaging.  The advantage of performing many
runs is that errors can be estimated (approximately) by taking the standard
deviation.  The present work represents one of the very few reports of
finite-size scaling of MC data which includes error bars.  All histograms
were generated at Kawamura's estimate of the critical temperature,
$T_c=1.458$.

The correlation time $\tau$ for the spin order parameter
was estimated \cite{tang} to be about 620 MCS at $L=24$ and $T=1.458$.
With the assumption $\tau \sim L^2$,
it can thus be expected that averaging was performed using
roughly 500 independent
configurations \cite{flb} in a single run for our largest lattice size.
Although $\tau$ decreases sharply away from $T_c$, it remains rather large.
At $T=1.440$, for example, $\tau$ was found to be approximately 200 MCS.
This result implies that averaging was made using not more than
about 8 independent configurations in a single
run for Kawamura's simulations at this temperature with $L=60$.

Results of applying the cumulant-crossing method\cite{pec2} to estimate the
critical temperatures associated with both spin and chiral orderings
are presented in Fig. 1.  The points represent the temperatures at which
the order-parameter cumulant $U_m(T)$ at $L'$ crosses the cumulant at
$L=12$ or $L=15$.
There is considerable scatter in the data and care must be taken to
use only results with $L$ sufficiently large to be in the
asymptotic region where a linear extrapolation is justified \cite{pec2}.
In the
case of the spin order, this appears to be for $Ln^{-1}(L'/L)$
\hbox{$<$\kern -0.8em\lower 0.8ex\hbox{$\sim$}} 1.5
but somewhat larger in the case of chiral order.  As with the Heisenberg
model, finite-size effects appear much less pronounced for the chiral
degrees of freedom \cite{mail1}.  These results suggest that the
two types of order occur at the same temperature, $T_c=1.4584(6)$.
The possibility that there are two {\it very} close but distinct
ordering temperatures can never be ruled-out based on finite-size simulations.

The possibility that the transition is weakly first order was also
examined.  No evidence for a double-peak structures was found in the
energy histograms, consistent with a continuous transition.  In addition,
the fourth-order energy cumulant $U$ evaluated at $T_c$ yielded
a result extrapolated to $L \rightarrow \infty$ of $U^\ast = 0.666~652(20)$,
consistent with a value $\frac23$ expected for a continuous transition
\cite{challa}.  A somewhat smaller value occurs in the case of a
weak first-order transition \cite{plum1}.  We note, however, that other
continuous transitions which occur in this model under the influence of an
applied magnetic field have values of $U^\ast$ closer to $\frac23$
than found in the present case \cite{plum1}.
Finally, the assumption of
volume-dependent scaling of various thermodynamic quantities did not yield
a straight-line fit, even for the data at large L.

Finite-size scaling results at $T_c$ for the specific heat $C$, spin order
parameter
$M$, susceptibility (as defined in Kawamura's work \cite{kawa2,mail1,pec2})
$\chi$, and the first logarithmic derivative of the order parameter
$V_1 = \partial [Ln(M)]/ \partial K$ (where $K=T^{-1}$)
\cite{ferr} are shown in Figs. 2-4.
Exponent ratios were estimated by performing Ln-Ln plots and also by
assuming a scaling dependence $F=aL^x$ for a function of interest
(plus a constant term in the case of the specific heat).
Except for the specific heat where errors are very large,
the two methods gave essentially the same results only if the smaller
lattices $L=12$ and $L=15$ were excluded.  In order to estimate errors
due to the uncertainty in $T_c$, identical scaling was also performed
at $T=1.4579$ and $T=1.4590$.   The resulting exponent ratios, along with
those associated with chiral order, are presented in Table I.
All results correspond to fits performed on data for $L=18-33$, except
in the case of the specific heat where $L=15$ data was also included
to reduce the error (otherwise, the result is $\alpha = 0.47(20)$).
The given
errors represent the robustness of the fitting procedure and do not
account for error bars on the figures.  Results for the exponents
$\nu$ and $\nu_\kappa$ estimated from the {\it second} logarithmic
derivative $V_2$ \cite{mail1,ferr} are 0.51(1) and 0.55(1),
respectively.

For ease of comparison, the
results of Kawamura's work are also included in the table.
Note that in order to obtain best-fit exponents
by the data-collapsing method, Kawamura used two different values of $T_c$:
1.458 for spin order and
1.459 for chiral order (within the range of our estimate
for $T_c$).  Within errors, however, he concludes
that the two transitions were the same.

{}From the results in Table I, it can be seen that $\beta, \gamma,
\beta_\kappa,$
and $\gamma_\kappa$ are the most sensitive
to the choice of $T_c$.  Variation in the exponents due to the
error bars on the figures was about the same.  With these considerations,
our best estimate of the exponents and their associated errors at
$T_c=1.4584(6)$
are given by: $\alpha = 0.46(10), \beta = 0.24(2), \gamma = 1.03(4),
\nu = 0.50(1)$ for spin order, and
$\beta_\kappa = 0.38(2), \gamma_\kappa = 0.90(9),
\nu_\kappa = 0.55(1)$ for chiral order.
These results for the spin order strongly suggest that
the transition is mean-field tricritial.

A possible interpretation for
this behavior can be found in our recent results of applying an in-plane
magnetic field \cite{plum1}.  This work reveals that $T_c$ is a multicritical
point, with one phase having the symmetry of the three-state Potts model and
consequently a weak first-order transition.  The effect of the field is
to generate a term third-order in ${\bf S}$ in the free energy,
of the form $\sim mS^3$,
where $m$ is the $q=0$ Fourier component of the spin density induced
by the field.  Since the system is frustrated in the triangular plane,
one might expect that short-range order along the $c$-axis is already
well developed at temperatures near $T_c$, in the present case corresponding
to the spin component $m$.  At zero applied field,
coupling between this short-range order and
$S$ could generate a third-order term which influences the critical
behavior.  For other types of systems, there may be similar coupling
to other Fourier components \cite{olle}.

It is also noteworthy that even though our simulations on the
XY model were made with about a factor of ten more MCS than in the
Heisenberg system \cite{mail1}, larger fluctuations were observed in the
present case for the spin order.
This can be observed by comparing the cumulant-crossing data
of Fig. 1 with the corresponding results of Ref.\cite{mail1}.
Larger fluctuations are expected if the transition is at or near a
tricritical point.

Our results for the chiral order are more difficult to interpret.
The values for the exponents are not too different from those of
Kawamura (when the errors are accounted for), and do not correspond to
any known universality class (also see Ref. \cite{chiral}).  The estimates for
$\nu_\kappa$ are in agreement but are significantly different from our
value of $\nu$ for the spin order.  This may indicate that the chiral order
has a distinct correlation length and that its criticality is
decoupled from the spin order
\cite{chiral,azapri}. Another possibility \cite{gilpri} currently
under investigation is that chirality
is associated with one of many possible symmetry-breaking fields,
in a manner analogous to the infinite set of crossover exponents previously
determined for the XY model \cite{ahar}.

In conclusion, these results of extensive histogram MC simulations
of the stacked triangular XY antiferromagnet strongly support the
scenario of tricritical behavior associated with the spin order,
in agreement with Azaria {\it et al.}
and in contrast with the proposal of new XY-chiral universality by Kawamura.
(Recent MC simulations do support, however, the existence of a
new Heisenberg-chiral universality class \cite{bhat,loi,mail1}.)
This resolves a long-standing controversy in the litterature, and
is relevant to a wide class of frustrated systems.
Further work is necessary to fully understand the nature of the
chiral ordering transition which appears to be decoupled from the
spin order.

\acknowledgements
We thank P. Azaria, H. Kawamura and G. Zumbach for useful discussions
and the Service de l'Informatique for the partial use of 10 RISC 6000
Workstations.  This work was supported by NSERC of
Canada and FCAR du Qu\'ebec.
%


\vfill\eject
\begin{table}
\caption{ Varation of exponents with assumed critical temperature,
along with Kawamura's results.}
\vskip0.2cm
\begin{tabular}{ccccc}
$T_c$            & 1.4579  & 1.4584   & 1.4590   & Kawamura \\
\tableline
$\alpha$         & 0.47(8) & 0.46(10) & 0.39(13) & 0.34(6)  \\
$\beta$          & 0.23(1) & 0.24(1)  & 0.26(1)  & 0.25(1)  \\
$\gamma$         & 1.07(2) & 1.03(2)  & 0.99(2)  & 1.13(5)  \\
$\nu$            & 0.51(1) & 0.50(1)  & 0.50(1)  & 0.54(2)  \\
$\beta_\kappa$   & 0.36(1) & 0.38(1)  & 0.40(1)  & 0.45(2)  \\
$\gamma_\kappa$  & 0.96(2) & 0.90(2)  & 0.81(2)  & 0.77(5)  \\
$\nu_\kappa$     & 0.56(1) & 0.55(1)  & 0.54(1)  & 0.55(2)  \\
\end{tabular}
\end{table}


\begin{figure}
\caption{Results of applying the cumulant-crossing method (see text)
to estimate the
critical temperatures associated with the spin and chiral orderings
where $b=L'/L$.}
\label{fig1}
\end{figure}

\begin{figure}
\caption{Finite-size scaling of the specific heat data for $L=12-33$.
Data at L=12 is excluded from the fit.  Error bars are estimated
from the standard deviation found in the MC runs.}
\label{fig2}
\end{figure}

\begin{figure}
\caption{Finite-size scaling of the order parameter.
Data at L=12 and 15 are excluded from the fit.  Error bars are estimated
from the standard deviation found in the MC runs.}
\label{fig3}
\end{figure}

\begin{figure}
\caption{Finite-size scaling of the susceptibility $\chi$
as well as the logarithmic derivative of the order
parameter $V_1$ (see text), as in Fig. 2.}
\label{fig4}
\end{figure}

\end{document}